\documentclass[aps,pre,twocolumn,groupedaddress,nofootinbib,notitlepage,showpacs,floatfix]{revtex4-1}
\usepackage{graphicx,graphics,epsfig,subfigure,times,bm,bbm,amssymb,amsmath,amsfonts,amsthm,mathrsfs,MnSymbol}
\usepackage[matrix,frame,arrow]{xypic}
\usepackage[pdfstartview=FitH]{hyperref}
\usepackage{pifont}
\hypersetup{
    colorlinks=true,       
    linkcolor=red,          
   citecolor=magenta,        
    filecolor=magenta,      
    urlcolor=cyan,           
    runcolor=cyan
}
\usepackage[pdftex]{color}
\usepackage{braket}
\usepackage{enumerate}
\usepackage[normalem]{ulem}
\usepackage{subfigure}
\usepackage{overpic}
\usepackage{dsfont}
\usepackage{geometry}
\usepackage{multirow}
\newcommand{\be}{\begin{equation}}
\newcommand{\ee}{\end{equation}}
\newcommand{\ba}{\begin{eqnarray}}
\newcommand{\ea}{\end{eqnarray}}

\newcommand{\ignore}[1]{}

\begin{document}

\title{Long-Range Order and Quantum Criticality in Antiferromagnetic Chains with Long-Range Staggered Interactions}

\author{Jie Ren}
\affiliation{Department of Physics, Changshu Institute of Technology, Changshu 215500, China}

\author{Zhao Wang}
\affiliation{Department of Physics, Changshu Institute of Technology, Changshu 215500, China}

\author{Weixia Chen}
\affiliation{Department of Physics, Changshu Institute of Technology, Changshu 215500, China}

\author{Wen-Long You}
\email{wlyou@nuaa.edu.cn}
\affiliation{College of Science, Nanjing University of Aeronautics and Astronautics, Nanjing, 211106, China}
\affiliation{MIIT Key Laboratory of Aerospace Information Materials and Physics, Nanjing University of Aeronautics and Astronautics, Nanjing 211106, China}

\date{\today}
\begin{abstract}
We study quantum phase transitions in Heisenberg antiferromagnetic chains with a staggered power-law decaying long-range
interactions. Employing the
density-matrix renormalization group (DMRG) algorithm and the fidelity susceptibility as the criticality measure, we establish more accurate values of quantum critical points than the results obtained from the spin-wave approximation, quantum Monte Carlo and DMRG in literatures. The 
deviation
is especially evident for strong long-range interactions. We extend isotropic long-range interactions to the anisotropic cases and find that kaleidoscope of quantum phases emerge from the interplay of anisotropy of the long-range exchange interaction and symmetry breaking. We demonstrate nonfrustrating long-range interactions induce the true long-range order in Heisenberg antiferromagnetic chains with a continuous symmetry breaking, lifting the restrictions imposed by the Mermin-Wagner theorem.
\end{abstract}
\maketitle

\section{Introduction}
As a prototypical model of magnetism, antiferromagnetic (AFM) Heisenberg model $H$= $\sum_{i,j}J_{i,j}\vec{S}_i\cdot\vec{S}_{j}$
has been persistently investigated for decades~\cite{Auerbach}.
Despite being a simplified theoretical model, the Heisenberg model finds applications in a variety of contexts, ranging from quantum phase transitions (QPTs)~\cite{Chen07,yi2019,Ren20,You20a,You22}, superconductivity~\cite{Mattis},  localization in disordered
systems~\cite{Znidaric}, spin liquid~\cite{Mei17}, quantum chaos~\cite{Kharkov20} to
quantum information~\cite{Nielsen00}.
The ground state of the nearest-neighbor AFM Heisenberg model on a bipartite lattice in $d$ ($\ge$ 2) dimensions is generally expected to host N\'{e}el long-range order (LRO) for any spin magnitude $S$, although a rigorous proof of the existence of LRO in a two-dimensional quantum-spin-1/2 Heisenberg magnet is still lacking~\cite{Tang89,Lin92,You09}.
It was recognized that imposed by the Mermin-Wagner theorem, the true LRO is prohibited in short-range interacting Heisenberg model in one spatial dimension.  Pioneering work by Haldane demonstrated that Heisenberg AFM chains of integer spins are endowed with a symmetry-protected topological gapped ground state~\cite{Haldane,Haldane1}, in stark contrast to the well-known spin-1/2 analog, which supports a quasi-long-range ordered critical phase, known as the
Tomonaga-Luttinger liquid (TLL). In this regard, higher dimensional magnets provide a testbed for spin-wave theory, while
the spin-wave approximation usually fails in one dimension. The remarkable
difference between one-dimensional (1D) AFM
systems of integer and half-integer spins opens a highly successful avenue in understanding the low-dimensional strongly correlated electronic materials. The isotropic Heisenberg AFM model has been unexpectedly coined 
in a number of nearly ideal quasi-one-dimensional materials such as Cu(C$_4$H$_4$N$_2$)(NO$_3$)$_2$~\cite{PhysRevB.59.1008}, Sr$_2$Cu(PO$_4$)$_2$~\cite{PhysRevB.74.174435}, KCuF$_3$~\cite{PhysRevLett.111.137205}, CuSO$_4\cdot 5$D$_2$O~\cite{Mourigal13}, and  spin-1 chain materials like SrNi$_2$V$_2$O$_8$~\cite{Bera13,Bera15}, Ni(C$_2$H$_8$N$_2$)$_2$NO$_2$(ClO$_4$)~\cite{Delica91,Avenel92} and NiI$_2$ (C$_7$H$_9$N)$_4$~\cite{Williams20}.
There have also been attempts to realize spontaneous symmetry
breaking and develop true AFM order in spin-1/2 Heisenberg
chains. One scheme under the consideration is the inclusion of the long-range interactions~\cite{defenu2021longrange}, which effectively increases the dimensionality
and lifts the rigorous restrictions imposed by the Mermin-Wagner theorem.

In fact, long-range interactions occur naturally in numerous quantum materials~\cite{Guo,Bramwell,Castelnovo,Annica} and versatile quantum simulators~\cite{Bloch1,Kim,Bloch2,Senko2015}.
Especially it has been suggested that the existing cavity-mediated cold atom system~\cite{Esslinger13} or  Rydberg dressed atoms~\cite{Saffman,Huse,Sde,Zoller2015} could be more ideal experimental platforms for long-range interactions than solid-state
ones.
For instance, the interacting radius of the effective interaction  between dressed atoms and the potential shape can be finely tuned by dressing to different fine-structure split states~\cite{Bouchoule,Zoller,Pohl,Li2012}.
The typical models
have considered interactions decaying with distance $r$ as a power law $\propto$ $1/r^\alpha$ or 
a staggered power law $\propto$ $(-1)^r/r^\alpha$,
ranging from dipolar spin chain~\cite{Noack}, Haldane-Shastry chain~\cite{Haldane88} to spin-1 chain~\cite{gong2016,gong20161}.  %
The effective exchange interactions mediated by either photons or Rydberg dressing are generally U(1) or $\mathbb{Z}_2$-symmetric, and a high degree of
symmetry, ideally SU(2), can be achieved by adjusting the laser detunings or increasing bosonic modes.  To be specific, it is found that the long-range interactions of the  
longitudinal component results in a Wigner crystal phase~\cite{Ren,Li2019}, whereas the transversal one may break a continuous symmetry, resulting in a continuous symmetry-breaking phase~\cite{Ren,Maghrebi}.

Inspired by the rapid development of quantum information science,   various information measurements have been exploited to study of quantum critical phenomena in spin chains.
The well-known and widely studied measures are entanglement entropy (EE)~\cite{Amico, Laflorencie} and fidelity susceptibility (FS), which diverges at the critical points in the thermodynamic limit~\cite{You,Gu2010}. The ground-state 
EE and FS were deemed to be capable of qualifying QPTs in many-body systems with short-range interactions~\cite{Ren2015,Nielsen,Ren2018,Bwang,Sun2015,Luo,Lv2022}, even for long-range 
interacting system~\cite{Ren2020,Sun2017,Sun2018}.  In the paper, we will detect the phase transitions in AFM Heisenberg chain with long-range anisotropic
interactions by the FS and the EE.

The remainder of this paper is organized as follows. We introduce the $S=1/2$ Heisenberg model with long-range anisotropic interactions in Sec. \ref{SEC:Hamiltonian}. The details of numerical methods and measurements are  also introduced. In Sec. \ref{SEC:Results}, effects of long-range interactions on correlation functions,
the FS and the EE are investigated. The discussion and summary are presented in the last Section.

\section{Hamiltonian and Measurements }\label{SEC:Hamiltonian}

In what follows, we are interested in a 1D spin-$1/2$ nearest-neighbor isotropic AFM chain under the effect of anisotropic long-range Heisenberg anisotropic interactions, given by
\begin{eqnarray}
\label{Hamiltonian}
H\!=\!J\!  \sum_{i}\left\{\!\vec{S}_i\cdot\vec{S}_{i+1}\!-\!\sum_{r\ge2}\!\lambda_{i,i+r} [ \Delta^{xy}\!(S^x_iS^x_{i+r}\!+\!S^y_iS^y_{i+r})
 \!+\!S^z_iS^z_{i+r}]\right\}, \nonumber \\
\end{eqnarray}
where $S_i^{\beta}$ ($\beta$ =$x$, $y$, $z$) are spin-1/2 
operators at $i$-th site among total $L$ sites.
The AFM coupling 
 $J$ = 1 between the nearest-neighbor spins is set up as an energy unit for simplicity unless otherwise stated. The connectivity  between two spins at sites $i$ and $i+r$ separated by a distance of $r (\ge 2)$ is given by
\begin{eqnarray}
\label{Jr}
\lambda_{i,i+r}\equiv \lambda(-1)^r r^{-\alpha},
\end{eqnarray}
for the nonfrustrating long-range interactions. Here we always choose the nearest-neighbor interactions to be isotropic, favoring a TLL ground state. The deformation of beyond-nearest-neighbour
couplings breaking from SU(2) symmetry down to a U(1) symmetry is characterized by the anisotropy parameter $\Delta^{xy}$ within the $x-y$ plane, which recovers isotropic interactions for $\Delta^{ xy}=1$ and reduces to $\mathbb{Z}_2$-symmetric Ising interactions for $\Delta^{xy}=0$. In this vein, the interplay of nearest-neighbor isotropic and longer-range anisotropic interactions admits certain magnetic symmetry breaking and stabilizes kaleidoscope of quantum phases. While
a solid-state implementation of Hamiltonian Eq.(\ref{Hamiltonian}) remains challenging,
engineering such graph of interactions can be possibly realized in state-of-the-art cavity QED~\cite{Chiocchetta21}.
To be explicit,
in an array of atomic ensembles within an optical cavity,
the strength of spin-spin interaction patterns, including the flip-flop and diagonal interactions as well as the decay exponent,
can be delicately resolved in a multimode cavity QED with an additional drive field~\cite{Vaidya18}, and the changing sign is determined by the phase of the corresponding sinusoidal modulation~\cite{Periwal21}. The unprecedented controllability of the cavity QED highlight the graph of interactions in Hamiltonian Eq.(\ref{Hamiltonian}) becomes programmable.

The simultaneous appearance of long-range interactions and symmetry breaking leads to quantum critical phenomena that is different from short-range interactions.
In the limit of $\alpha\rightarrow \infty$ or $\lambda \rightarrow 0$, the Hamiltonian in Eq. (\ref{Hamiltonian}) is reduced to a spin-1/2 chain solely with the nearest-neighbor interactions, which can be analytically solved by Bethe ansatz~\cite{Bethe31}.  For generic parameters $\{\alpha, \lambda\}$,
the system becomes nonintegrable.
It is anticipated that the ground state is still in a quasi-long-range ordered phase for a sufficiently large $\alpha$, while the
system favors long-range order for a small value of $\alpha$. For the 
Heisenberg chain with staggered power-law decaying interactions, the transition between LRO phase and quasi-long-range order (QLRO) was successively investigated in literature~\cite{Parreira,Affleck,Sandvik,Yang2020}.  At first Parreira {\it et al.} pointed out that
the LRO is absent for $\alpha>3$ with any $\lambda$ based on spin-wave theory~\cite{Parreira}, indicating that the critical line $\alpha_c<3$ between the AFM N\'{e}el order and the QLRO phase.
Lately Laflorencie {\it et al.}  utilized the staggered
structure factor as order parameter to detect the N\'{e}el instability in terms of quantum Monte Carlo (QMC) simulations.
For $\lambda=1$ and $\Delta^{xy}=1$, they obtained
 $\alpha_c^{\rm QMC} = 2.225$, which improved the numerical results $\alpha_c^{\rm SW}=2.46$ given by the lowest order spin-wave approximation~\cite{Affleck}. Recently Yang {\it et al.} studied the QPTs from the perspective of the fractionalized excitations for chains of length $L = 60$ using 400 density-matrix renormalization group (DMRG) states~\cite{Yang2020}.  The development of the LRO is associated with the formation of coherent magnons that emerge from deconfined spinons in the gapless Luttinger liquid, giving rise to $\alpha_c^{\rm DMRG}\approx2.2$. Thus, it would be interesting to identify the accurate value of the critical point across this unconventional phase transition by other observables with enhanced sensitivity.

As a quantum information metric, the FS has proved to be particularly useful for detecting the critical points of symmetry-knowledge unknown systems~\cite{Wang18,PhysRevA.97.013845,Liu21}. For a general many-body Hamiltonian $H(g)$,
the ground-state FS per site can be calculated 
by~\cite{Gu2010,You}
\begin{equation}
\label{FS1}
\chi(g)=\lim_{\delta g\rightarrow
0}\frac{-2 \textrm{ln}F(g,\delta g)}{L(\delta g)^2},
\end{equation}
where the fidelity $F$ measures the similarity between the two closest ground states $|\psi_0(g)\rangle$ and $|\psi_0(g+ \delta g)\rangle$, which is defined as
\begin{equation}
\label{F}
F(g,\delta g)=|\langle\psi_0(g)|\psi_0(g+ \delta g)\rangle|.
\end{equation}
Here $g$ is the variational parameter of $H(g)$ and $\delta g$ denotes an infinitesimal deviation.
Note that Hamiltonian (\ref{Hamiltonian}) can not be expressed as a simple form as $H(\alpha)=H_0+\alpha H_I$.
Subsequently, we obtain the derivatives of Eq.(\ref{Jr}) as $\delta \lambda_{i,i+r}=-\lambda (-1)^r r^{-\alpha}\ln r   \delta \alpha$.  Due to nonfrustrated
characteristics, the average derivatives of interactions per site is practically considered as an effective tuning parameter $\delta \bar{\alpha}= \sum_{i<j}\delta \lambda_{i,j}/L$. Therefore, the 
FS per site can be calculated numerically by
\begin{equation}
\label{FS2}
\chi(\alpha)=\lim_{\delta \alpha\rightarrow
0}\frac{-2 \textrm{ln}F(\alpha,\delta \alpha)}{L(\delta \bar{\alpha})^2}.
\end{equation}
The peak of FS per site is thus used to identify the phase boundary $\alpha_c$ for continuously varying parameters $\{ \lambda, \Delta^{xy}\}$, which provides a vital opportunity to testify theoretical predictions with experimentally accessible results. Another familiar probe to monitor critical point is the bipartite von Neumann EE, which is defined by
\begin{eqnarray}
\label{EE} S_A=-\textrm{Tr}(\rho_{A}\ln\rho_{A}).
\end{eqnarray}
Here $\rho_A$ is the reduced density matrix of subsystem $A$ with
respect to the whole system. The EE 
can also be extracted from the ground-state wavefunction $|\psi_0\rangle$
 and hence properly 
 characterize the 
 QPTs. The ground states of short ranged Hamiltonians usually satisfy an area law
according to which the EE $S_A$ of a subregion $A$ of the system is proportional
to the size of its boundary area. This area-law conjecture is
be derived from the power-law decay of the bipartite
correlations~\cite{Hastings06} and numerically verified in various quantum many-body systems, and
is expected to be true in all noncritical phases~\cite{Amico}, even for long-range interacting systems~\cite{Kuwahara20}.
However, a logarithmic violation of the area law is
usually known to occur in critical ground states, as is coined by conformal field theory (CFT), where the system size $L$ is related to the correlation length $\xi$ near the critical point such as $L \sim \xi$ and the gap decays as $1/L$. In this case,  a coefficient proportional to the central charge of the underlying
CFT, the half-chain EE of 1D critical
systems of finite size $L$ with open boundary condition satisfy
\begin{eqnarray}
\label{eq3} S_{\rm h}(L)= \frac{c}{6}\ln  L + S_0,
\end{eqnarray}
where $c$ is the central charge, and $S_0$ is a nonuniversal constant.
However, the area law for long-range
interacting systems is still elusive. The conformal symmetry will break down under the long-range interactions when $\alpha$ is small~\cite{Vodola2014,Vodola2016,Maghrebi}, as the long-range interactions results in correlation patterns
similar to those in critical phases. To this end,  we calculate the effective central charge $c_{\rm eff}$ as a function of $\alpha$, which is obtained by calculating the half-chain EE for two chains with different $L_1$ and $L_2$. By using finite-size DMRG algorithm,  the effective central charge can be obtained by
 \begin{equation}
c_{\rm eff}=\frac{6[S_{\rm h}( L_2 )-S_{\rm h}(L_1)]}{\ln(L_2)-\ln(L_1)}.
\label{eq10}
\end{equation}
We emphasize that $c_{\rm eff}$  may not have the meaning of the central charge for the short-range interacting cases with conformal symmetries, although we find the half-chain EE always obeys the  scaling form in (\ref{eq3}).

A precise numerical determination of $\alpha_c$ poses significant technical challenges in terms of various criticality measures.
Theoretically, the treatment of quantum many-body systems is notoriously complicated so that many investigations are still accessible by numerical techniques like
the DMRG method~\cite{white,KWHP,U01}, the present studies of Hamiltonian (\ref{Hamiltonian}) can be simulated with very high accuracy.  Based on matrix product states, we adopt both infinite-size DMRG (iDMRG)~\cite{McCulloch} and finite-size DMRG~\cite{U02} where up to $m=2000$ in the truncation of bases are kept, and this allows the truncation error to be smaller than $10^{-9}$. The long-range interactions can be approximated by a summation of finite exponential terms~\cite{Vidal,Pirvu}, which inevitably introduces additional systematic error and corrupts the numerical results of FS.
In our calculations of finite-size DMRG algorithm, we handle with the long-range interactions 
using a summation over matrix product operators (MPOs). Our codes are mainly based on iTensor C++ library~\cite{tensor}. Since the $z$-component of the total spins for the present system commutes with the Hamiltonian (\ref{Hamiltonian}), the ground-state energy is obtained by comparing the lowest energies for each subspace of $S^z_{\rm tot}=\sum_{i=1}^L \langle S^z_i\rangle$. The ground state resides in the sector $S^z_{\rm tot}=0$ as a consequence of the continuous U(1) symmetry therein.

\section{Results}\label{SEC:Results}
With the 
DMRG algorithm at hand, we analyze the kaleidoscope of quantum phases that emerge in this system for different types of long-range exchange interactions. In the following, we will consider isotropic ($\Delta^{xy}=1$), Ising-type ($\Delta^{xy}=0$) and  XY-type ($\Delta^{xy}=1.5$) anisotropic cases, respectively. Using the powerful tools, the phases of long-range interacting systems are numerically diagnosed and the corresponding phase diagrams are determined.

\subsection{$\Delta^{xy} =1$}
\begin{figure}[h]
\includegraphics[width=0.40\textwidth]{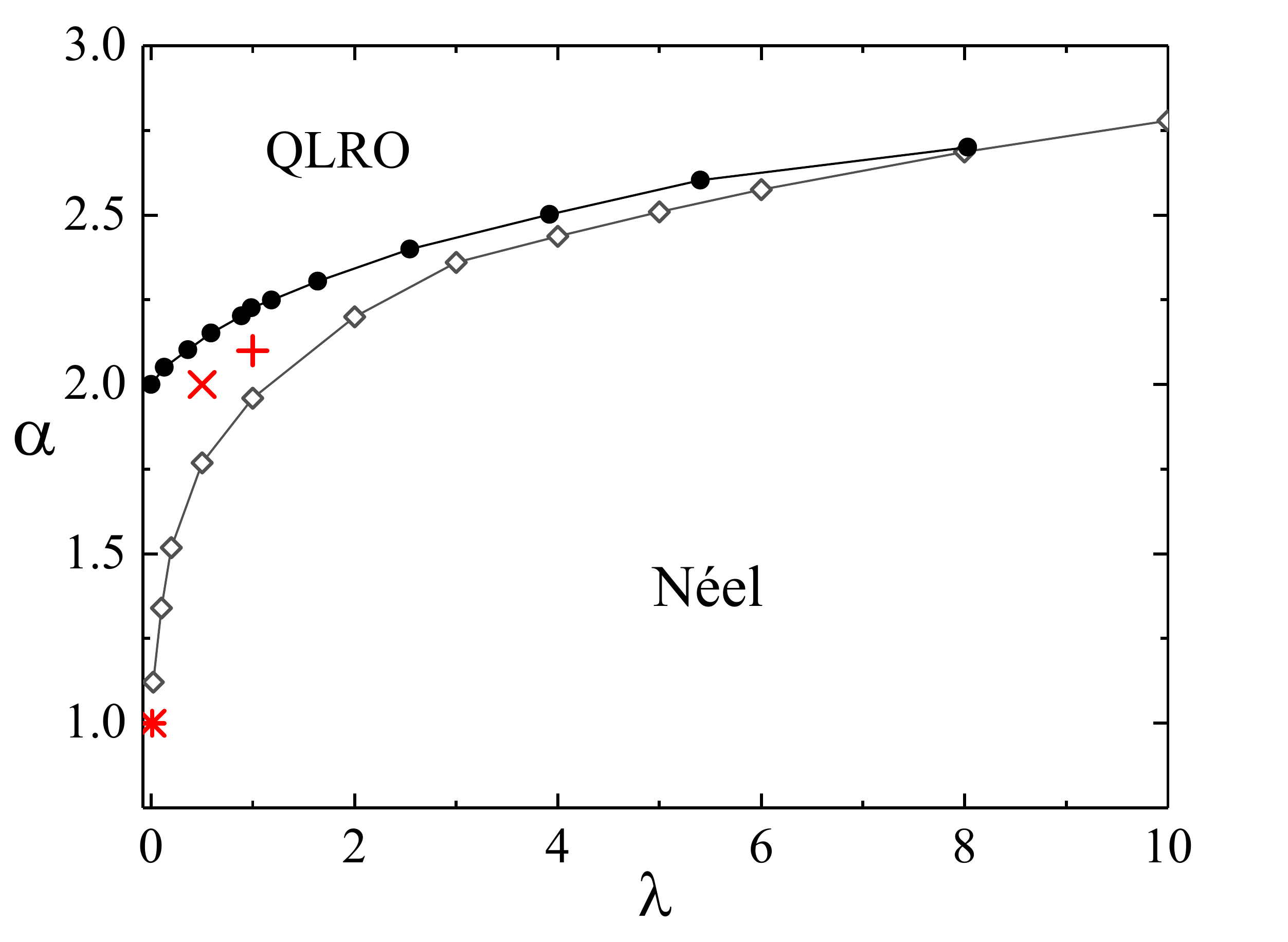}
\caption{\label{Phase_Heisenberg}Phase diagram of Hamiltonian Eq. (\ref{Hamiltonian}) as functions of $\alpha$ and $\lambda$ with $\Delta^{xy} =1$. The boundary (-$\bullet$-) between QLRO and LRO is computed by the large scale QMC simulation~\cite{Affleck}, and the results (-$\diamond$-) is obtained by the FS.  It is noted that the  LRO  phase is equivalent to  N\'{e}el phase in 1D spin systems. The symbols ($\ast$,$\times$,$+$) mark the positions of parameters used in Fig.\ref{correlation_H}.}
\end{figure}

\begin{figure}[h]
\includegraphics[width=0.40\textwidth]{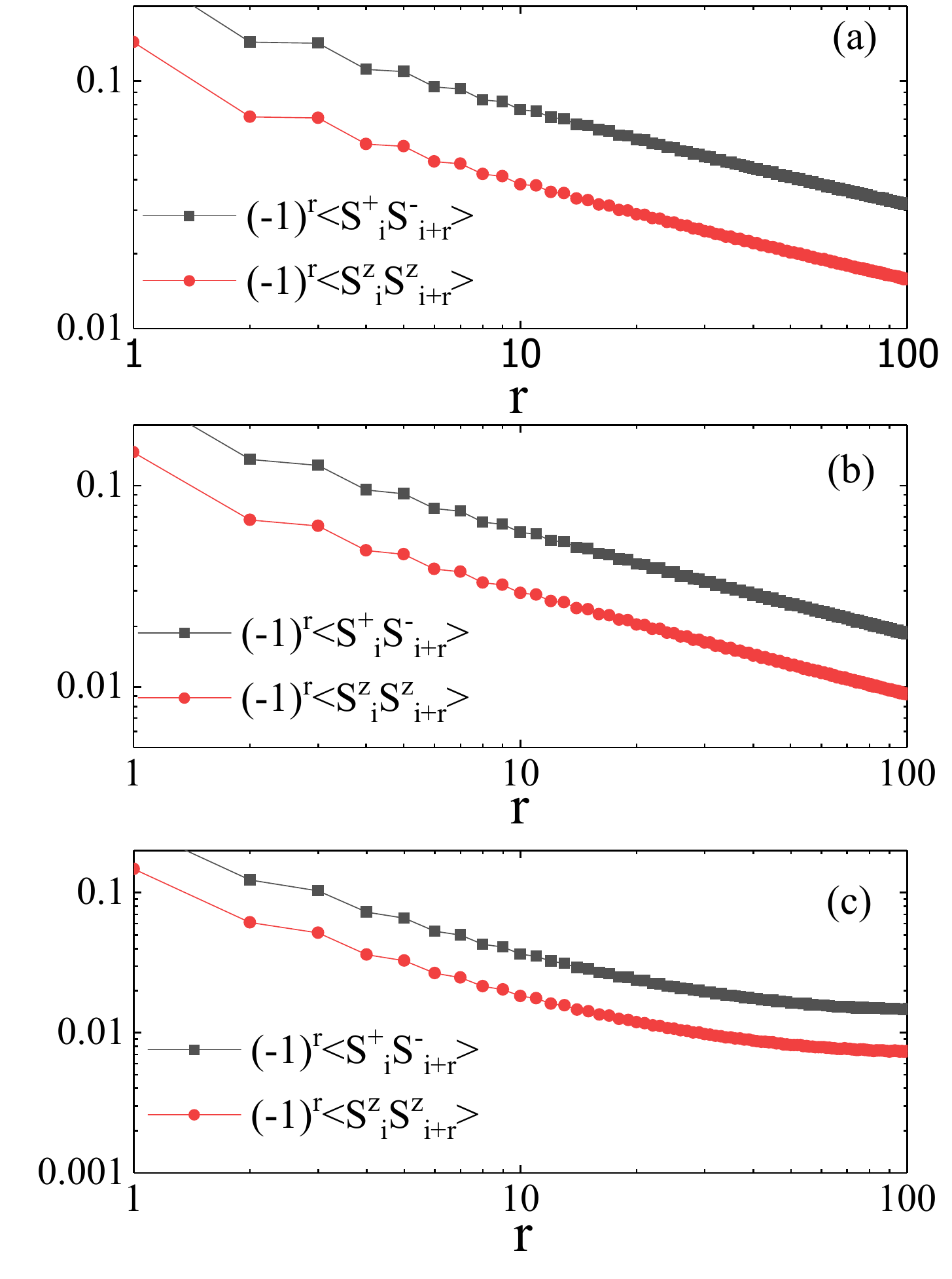}
\caption{\label{correlation_H} Loglog-plot correlations $\langle S^{+}_iS^{-}_{i+r}\rangle$, $\langle S^{z}_iS^{z}_{i+r}\rangle$ versus the distance
$r$ with $\Delta^{xy}=1$ for (a) $\lambda=1$, $\alpha=2.10$; (b) $\lambda=0.5$, $\alpha=2$ and (c) $\lambda=0.01$, $\alpha=1$.
}
\end{figure}

\begin{figure}[t]
\includegraphics[width=0.5\textwidth]{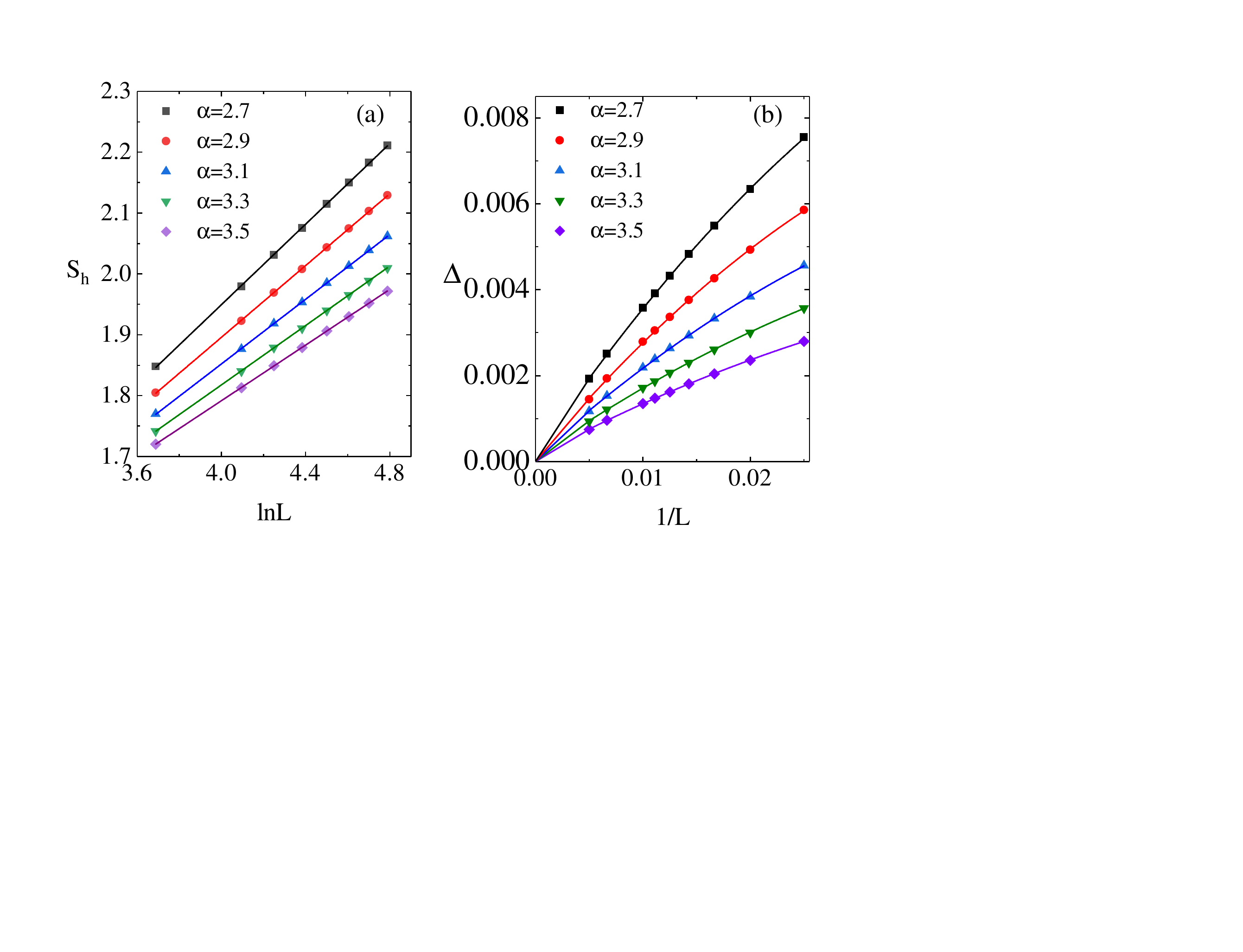}
\caption{(a) Half-chain EE versus $\ln L$ for different $\alpha$ with $\Delta^{xy} =1$, $\lambda=\infty$. Symbols show numerical results obtained by DMRG calculations, and solid lines are linear fits of the data. (b) Finite-size scaling of the energy gap $\Delta$ with various $\alpha$. Symbols show numerical results obtained by DMRG calculations, and solid lines are fits of the data by quadratic polynomials in 1/$L$.
\label{EE_scaling}}
\end{figure}

For the long-range
isotropic Heisenberg interactions, i.e., $\Delta^{xy} =1$, by using a combination of 
QMC and analytic methods, Laflorencie  {\it et al.}  have studied the phase diagram in the $\lambda$-$\alpha$ plane~\cite{Affleck}, 
as is shown in Fig.~\ref{Phase_Heisenberg}.  It is shown that the critical point 
between the N\'{e}el phase and the QLRO phase 
increases sharply from $\alpha^c(\lambda=0^+)=2$  to  $\alpha^c (\lambda=8)\approx 2.7$. To further understand
two phases, we investigate the correlation 
functions of the system using iDMRG, which can avoid the boundary effects. 
The correlation functions $\langle S^z_{i}S^z_{i+r}\rangle$ and $\langle S^+_{i}S^-_{i+r}\rangle$ with respect to the distance $r$ for $\alpha=2.1$, $\lambda=1$
are shown in Fig.~\ref{correlation_H}(a).
As we know, for 1D 
spin-1/2 short-range AFM Heisenberg system, the spin-spin correlation function
\begin{eqnarray}
\label{CC} \langle \vec{S}_{i}\cdot\vec{S}_{i+r}\rangle \propto \frac{(-1)^r\sqrt{\ln r}}{r},
\end{eqnarray}
is expected to characterize the QLRO phase, and
\begin{eqnarray}
\lim_{r\rightarrow +\infty}\langle \vec{S}_{i}\cdot\vec{S}_{i+r}\rangle=\pm m^2_{c}
\end{eqnarray}
is capable of identifying the N\'{e}el phase~\cite{Affleck1989}.
The power-law decay of $\langle S^z_{i}S^z_{i+r}\rangle$, $\langle S^+_{i}S^-_{i+r}\rangle$ in Fig.~\ref{correlation_H}(a), implies  $\lim_{r\rightarrow \infty}\langle \vec{S}_{i}\cdot\vec{S}_{i+r}\rangle$
would approach zero and the ground state for $\{\alpha=2.1$, $\lambda=1\}$ under consideration is within the QLRO phase and thus the critical point of N\'{e}el-to-QLRO transition should be below $2.1$ for $\lambda=1$.
The spatial correlation functions for $\{\alpha=2$, $\lambda=0.5\}$ are also calculated, and the QMC results showcase the system should be in
 N\'{e}el phase. One finds  
$\langle S^z_{i}S^z_{i+r}\rangle$ and $\langle S^+_{i}S^-_{i+r}\rangle$  also  
exhibit a power-law decay, as is observed in Fig.~\ref{correlation_H}(b), which means the 
ground state remains the QLRO phase.  
These above mentioned discoveries indicate the critical points retrieved by the QMC are not accurate. A more creditable measure should be adopted to determine the phase boundaries.
\begin{figure}[t]
\includegraphics[width=0.5\textwidth]{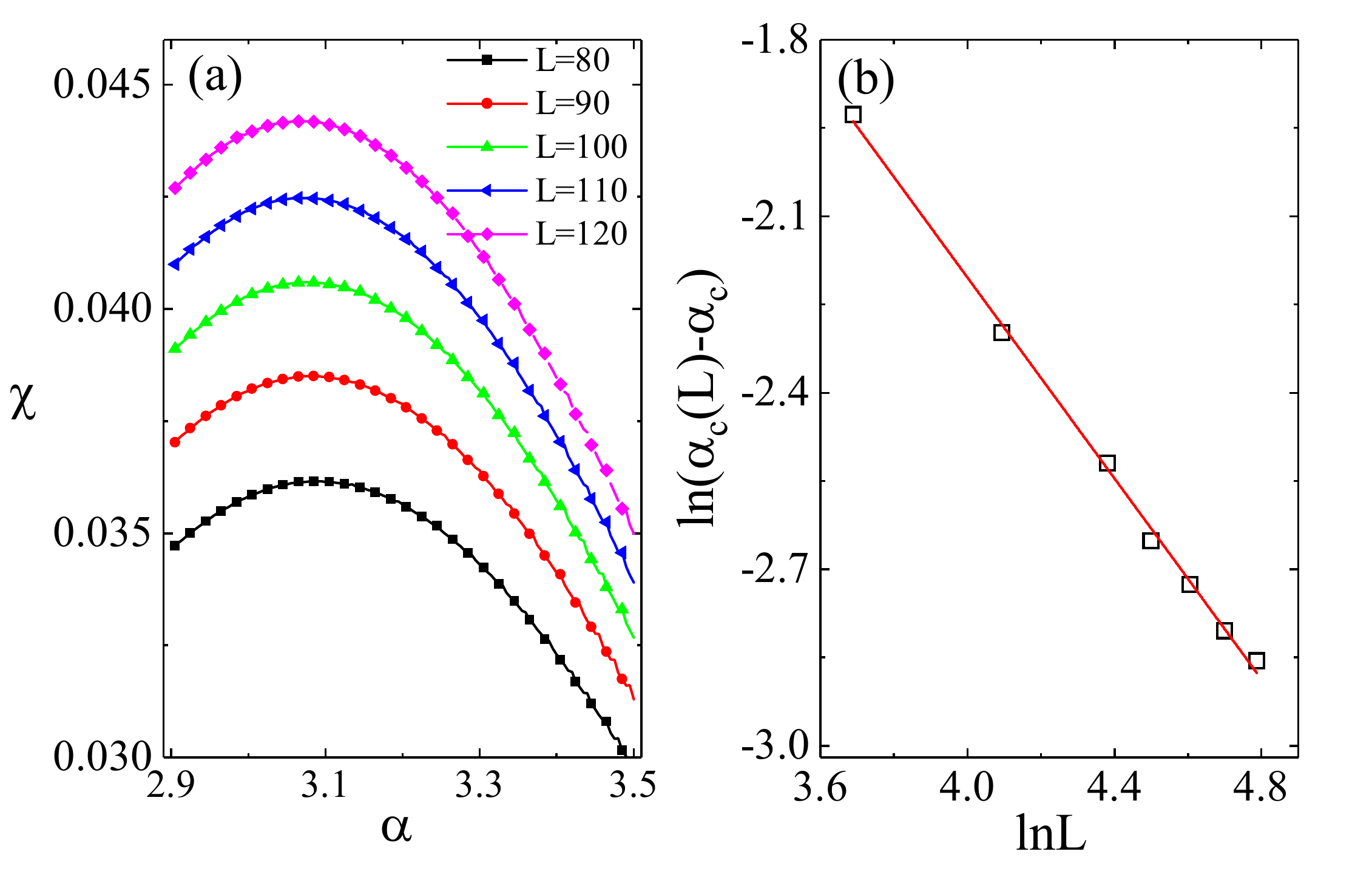}
\caption{(a) Fidelity susceptibility per site is plotted as a function of the parameter $\alpha$ on different system sizes $L$ with $\Delta^{xy} =1$, $\lambda$= $\infty$. (b) Scaling of the peak positions of $\chi$.
\label{FS0}}
\end{figure}

\begin{figure}[t]
\includegraphics[width=0.5\textwidth]{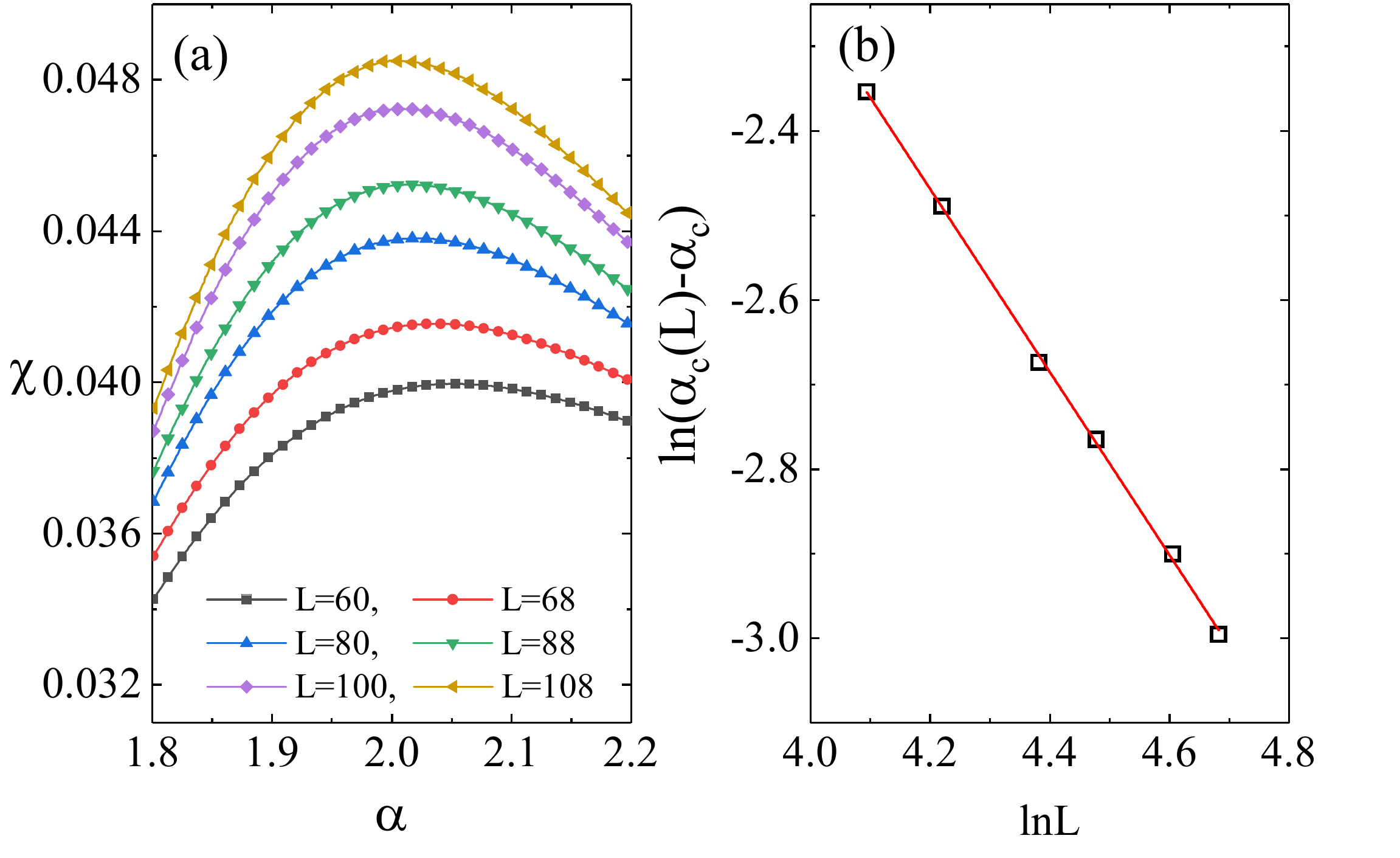}
\caption{(a) Fidelity susceptibility per site is plotted as a function of the parameter $\alpha$ on different system sizes $L$ with $\Delta^{xy} =1$, $\lambda=1$. (b) The corresponding scaling of the peak positions of $\chi$.
\label{FS_Heisenberg_J1}}
\end{figure}

To alleviate the controversy by the discrepancy between the QMC results (cf. Fig.\ref{Phase_Heisenberg}) and correlations (cf. Fig.\ref{correlation_H}),
we consider a limiting case, i.e., $\lambda$$\to$$\infty$, which can be equivalently implemented by switching off the nearest-neighbor isotropic interactions in Hamiltonian (\ref{Hamiltonian}) with finite $\lambda$.
The absence of the LRO has been rigorously proven for $\alpha >3$ with $\lambda=1$~\cite{Parreira,Roscilde},
and was lately extended to arbitrary $\lambda$~\cite{Affleck}. The critical point $\alpha_c^{\rm SW}$($\lambda$$\to$$\infty$)=2.9032 was inferred by the spin-wave approximation~\cite{Affleck}.
In this case, we use the EE to speculate the critical point.
We find the EE decreases monotonically with increasing $\alpha$.
In particular, the EE always shows a logarithmic growth with the system size as $S_{\rm h}\propto \ln L $ [Fig.~\ref{EE_scaling}(a)], 
which can be treated as reminiscent of gapless ground state in both the QLRO phase and 
the N\'{e}el phase~\cite{Yang2020}, 
as is indicated in Fig.~\ref{EE_scaling}(b).
Consequently the signal of the QPT is hardly discerned from the EE.

The impetus to identify the precise position of the quantum critical point(QCP) 
was given by the FS, which has been proven to be capable of detecting the phase transition successfully between two gapless phases
~\cite{Ren}. To this end, we will adopt the FS to identify the 
QCP between the N\'{e}el phase with LRO and the QLRO phase for $\lambda=\infty$ as a glimpse.  The numerical results are shown in Fig.~\ref{FS0}(a). One can observe the peak of the FS increases with the system size $L$ 
nearby $\alpha=3.1$. In order to locate the critical points $\alpha_c$ in the thermodynamic limit, we have used the finite-size scaling theory~\cite{Fisher}, which can be used in finite systems with long-range interactions~\cite{Loscar}. The position of the maximal points of the FS can be fitted by the following formula:
 \begin{equation}
|\alpha_{c}(L)-\alpha_c|\sim L^{-b},
\label{eq5}
\end{equation}
where $b$ is a constant and $\alpha_c$ is the QCP in the thermodynamic limit. For properly chosen values
of $\alpha_c=3.00$, $b=0.85$, we can see from Fig.~\ref{FS0}(b) that a linear relation following Eq.(\ref{eq5}) for different $L$ is verified.
Our results indicate the critical points 
 $\alpha_c $ would approach 
 3.0 as $\lambda\rightarrow \infty$. Recall that Parreira {\it et al.} pointed out the nonexistence of the N\'{e}el phase at zero temperature for $\alpha >3$ for $\lambda=1$~\cite{Parreira} and a straightforward extension 
 for all $\lambda$~\cite{Affleck}.
In this sense, 
 the surprising consistence between our result with the previous results
 confirm that the FS shows high accuracy and reliability in detecting the critical point of 
 the N\'{e}el-to-QLRO transitions.

Next we investigate the case of $\lambda=1$. The FS results for various system sizes are shown in Fig. \ref{FS_Heisenberg_J1}(a). The corresponding finite-size scaling according to Eq. (\ref{eq5}) is illustrated in Fig. \ref{FS_Heisenberg_J1}(b), giving rise to $\alpha_c=1.955$, $b=1.0$.
In contrast to the QMC result $\alpha_c^{\rm QMC}=2.225\pm0.025$ and the spin-wave result $\alpha_c^{\rm SW}=2.46$, the obtained value of $\alpha_c$ indicates that the ground state for $\alpha=2.1$ is within the QLRO phase. This is consistent with the correlations in Fig.~\ref{correlation_H}(a).
To this end, the FS is calculated for different $\lambda$ and the positions of critical points can be precisely retrieved from the FS results for $\lambda \ge 0.02$. One finds $\alpha_c\simeq 1.12$ for $\lambda=0.02$, whereas 
the positions of the critical points become elusive through the peak of the FS for $\lambda<0.02$.
As is observed in Fig.~\ref{correlation_H}(c), it is found that
the correlations $\langle S^z_{i}S^z_{i+r}\rangle$, $\langle S^+_{i}S^-_{i+r}\rangle$ tend to a constant for $\{$$\alpha=1$, $\lambda=0.01$$\}$,
implying that the critical point $\alpha_c \ge 1$  when $\lambda\rightarrow 0^+$, which is consistent with spin-wave result.
Based on the above analysis of correlation functions and the FS, we obtain the critical points and establish the ground-state phase diagram in Fig. \ref{Phase_Heisenberg}.  It is clear that the critical values $a_c(\lambda)$ get lower than those obtained by the large scale QMC simulation. The 
deviation is extremely prominent for small $\lambda$ but negligible for large $\lambda$.

\subsection{$\Delta^{xy} =1.5$}
\begin{figure}[h]
\includegraphics[width=0.43\textwidth]{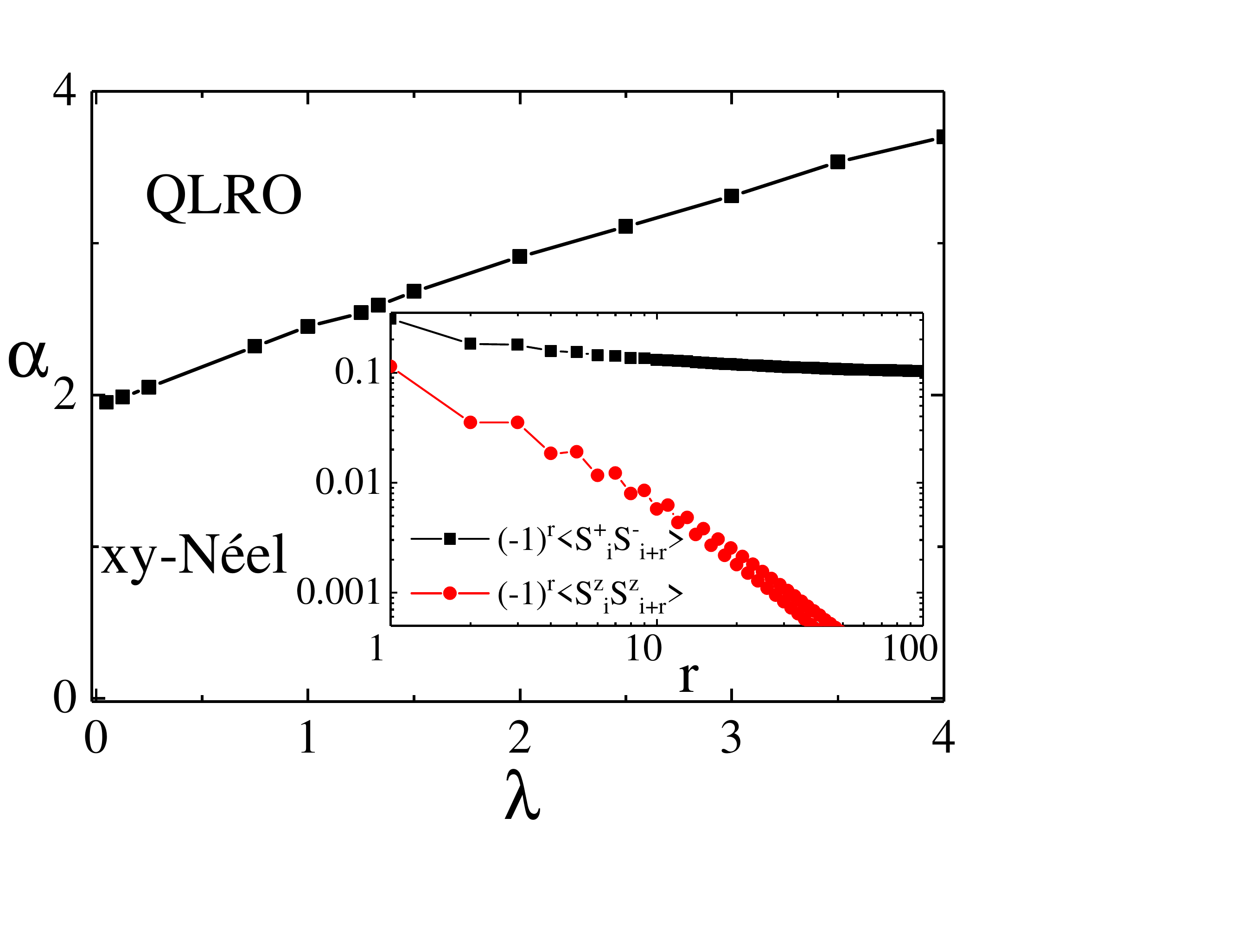}
\caption{\label{Phase_Jxy} Phase diagram of Hamiltonian Eq. (\ref{Hamiltonian}) as functions of $\alpha$ and $\lambda$ with $\Delta^{xy} =1.5$. Inset: Correlations $\langle S^{+}_iS^{-}_{i+r}\rangle$, $\langle S^{z}_iS^{z}_{i+r}\rangle$ versus the distance $r$ for $\lambda=1$, $\alpha=2$ with $\Delta^{xy} =1.5$.}
\end{figure}
\begin{figure}[t]
\includegraphics[width=0.425\textwidth]{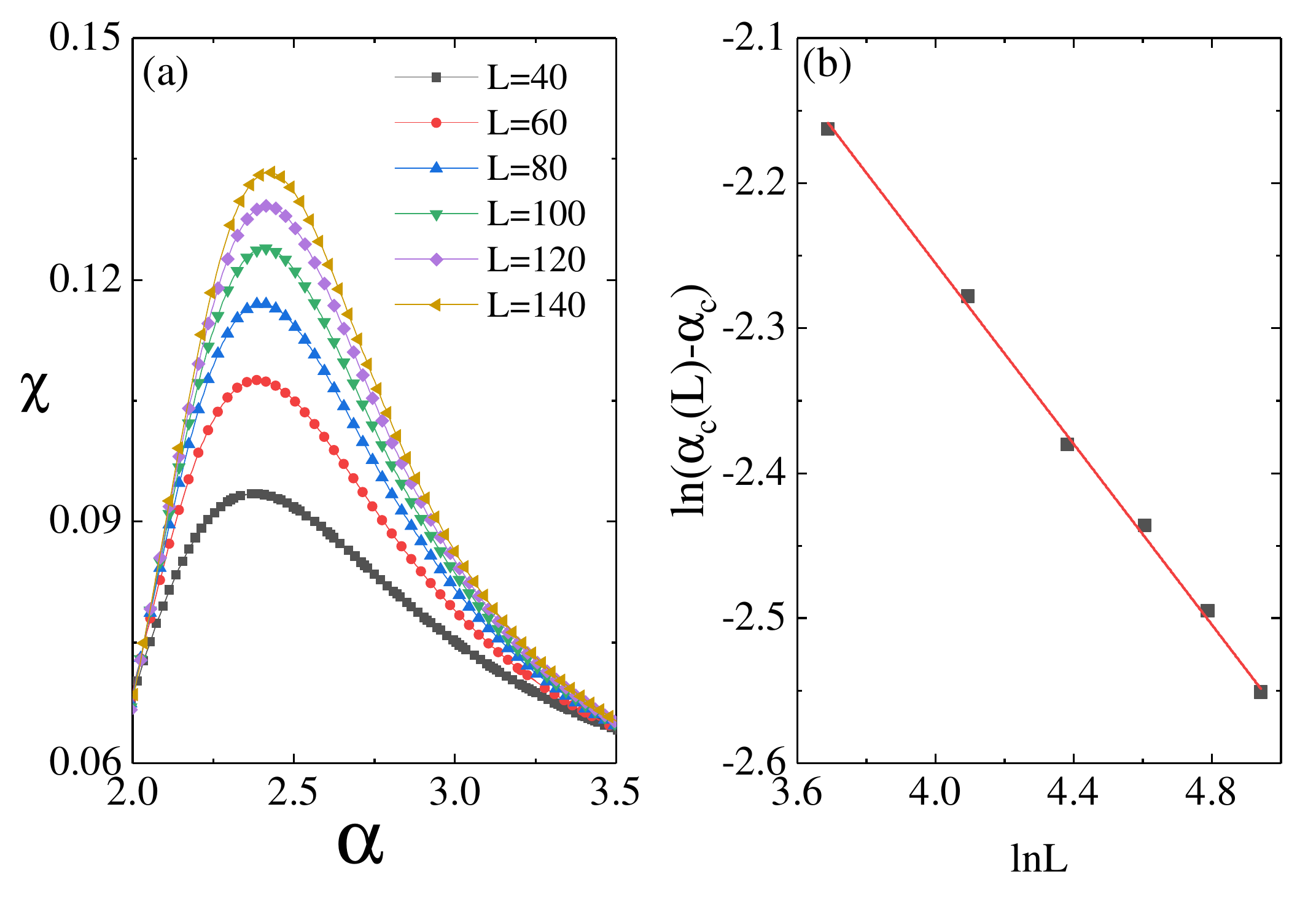}
\caption{(a) Fidelity susceptibility per site is plotted as a function of the parameter $\alpha$ on different system sizes $L$ with $\lambda=1$, $\Delta^{xy} =1.5$. (b) The corresponding scaling of the peak positions of $\chi$.
\label{FS_Jxy}}
\end{figure}

\begin{figure}[t]
\includegraphics[width=0.5\textwidth]{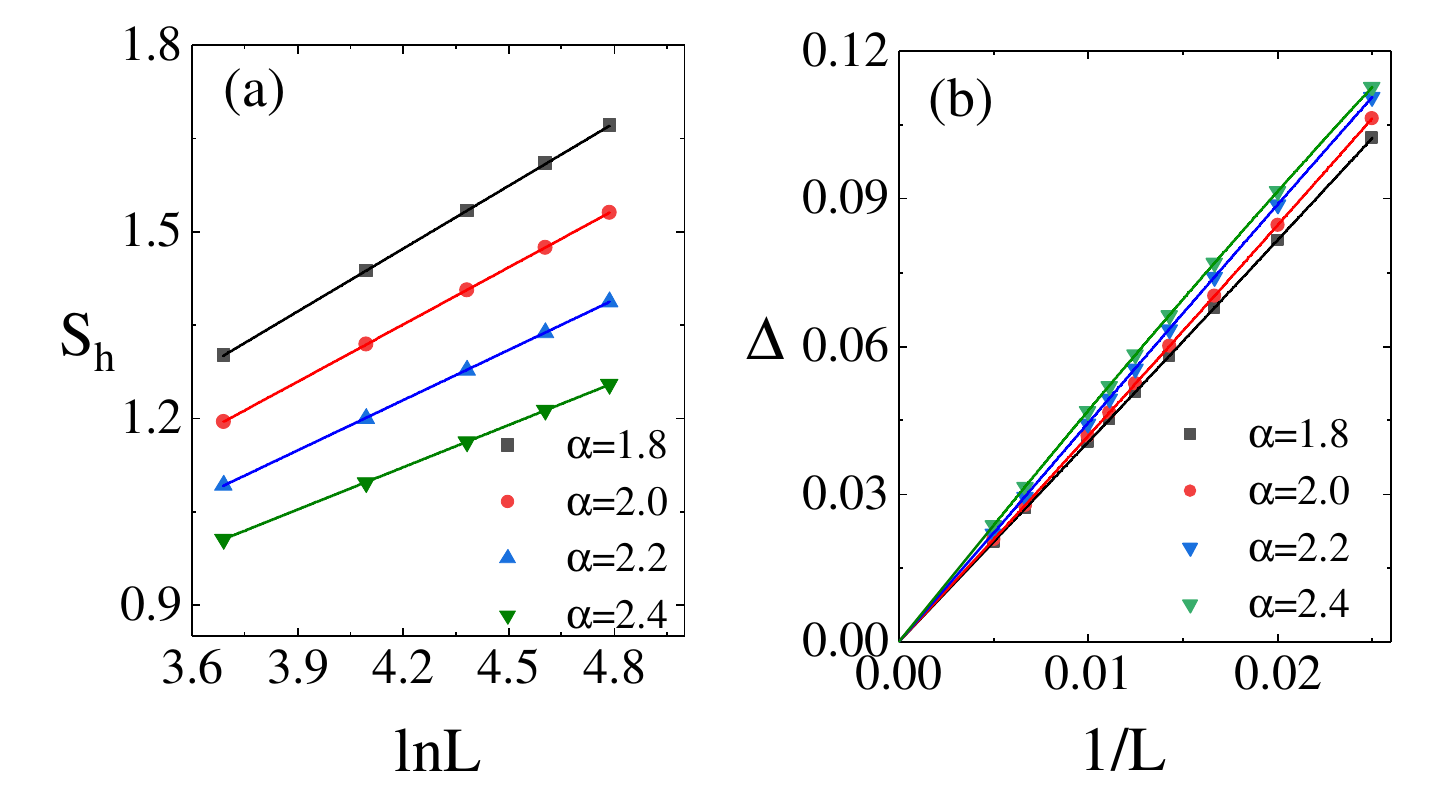}
\caption{(a) Half-chain EE versus $\ln L$ for different $\alpha$ with $\Delta^{xy} =1.5$, $\lambda=1$. (b) Finite-size scaling of the energy gap $\Delta$ with various $\alpha$. Symbols show numerical results obtained by DMRG calculations, and solid lines are linear fits of the data in 1/$L$. \label{EE_Jxy}}
\end{figure}

Next, we begin to study the effect of anisotropy of
long-range exchange interactions. First, the XY-type ($\Delta^{xy} >1$) exchange interactions are considered.
The phase diagram of Hamiltonian Eq. (\ref{Hamiltonian}) with $\Delta^{xy} =1.5$ as functions of $\alpha$ and $\lambda$ is shown in Fig.~\ref{Phase_Jxy}. For sufficiently large $\alpha$, the system would be in the QLRO phase. As the decay exponent $\alpha$ gets smaller, the long-range interactions will become dominated.
The correlation functions for $\{\alpha=2$, $\lambda=1\}$ 
are shown in the inset of Fig. \ref{Phase_Jxy},
where $\langle S^z_{i}S^z_{i+r}\rangle$ tends to vanish as $r \to \infty$, while $\langle S^+_{i}S^-_{i+r}\rangle$ will alternate between $-0.1$ and $0.1$, which means that 
the $xy$-N\'eel phase is stabilized with breaking of the
continuous U(1) symmetry in the $x-y$ plane. In a sense, the correlation function $\langle S^+_{i}S^-_{i+r}\rangle$ can act as an order parameter for the QPT between the QLRO and U(1)-symmetric broken phase. The ground-state FS per site $\chi$ for $\lambda=1$ is exhibited in Fig.~\ref{FS_Jxy}(a). Following the similar strategy as SU(2) symmetric model, the critical point $\alpha_c=2.42$ between the $xy$-N\'{e}el and QLRO phase is identified from Fig.~\ref{FS_Jxy}(b). Similarly,
the EE scales logarithmically with the system size in the $xy$-N\'{e}el phase, as is disclosed in Fig.~\ref{EE_Jxy}(a), suggesting that the $xy$-N\'{e}el phase remains gapless.
In order to validate the gapless nature,
the finite-size energy gap $\Delta(L)$ is calculated for  $\alpha<\alpha_c$ in Fig.~\ref{EE_Jxy}(b). The linear fitting with respect to $1/L$ designates that $\Delta(\infty)$ will vanish in the thermodynamic limit and the dynamical exponent $z=1$.

\subsection{$\Delta^{xy} =0$}
\begin{figure}[h]
\includegraphics[width=0.4\textwidth]{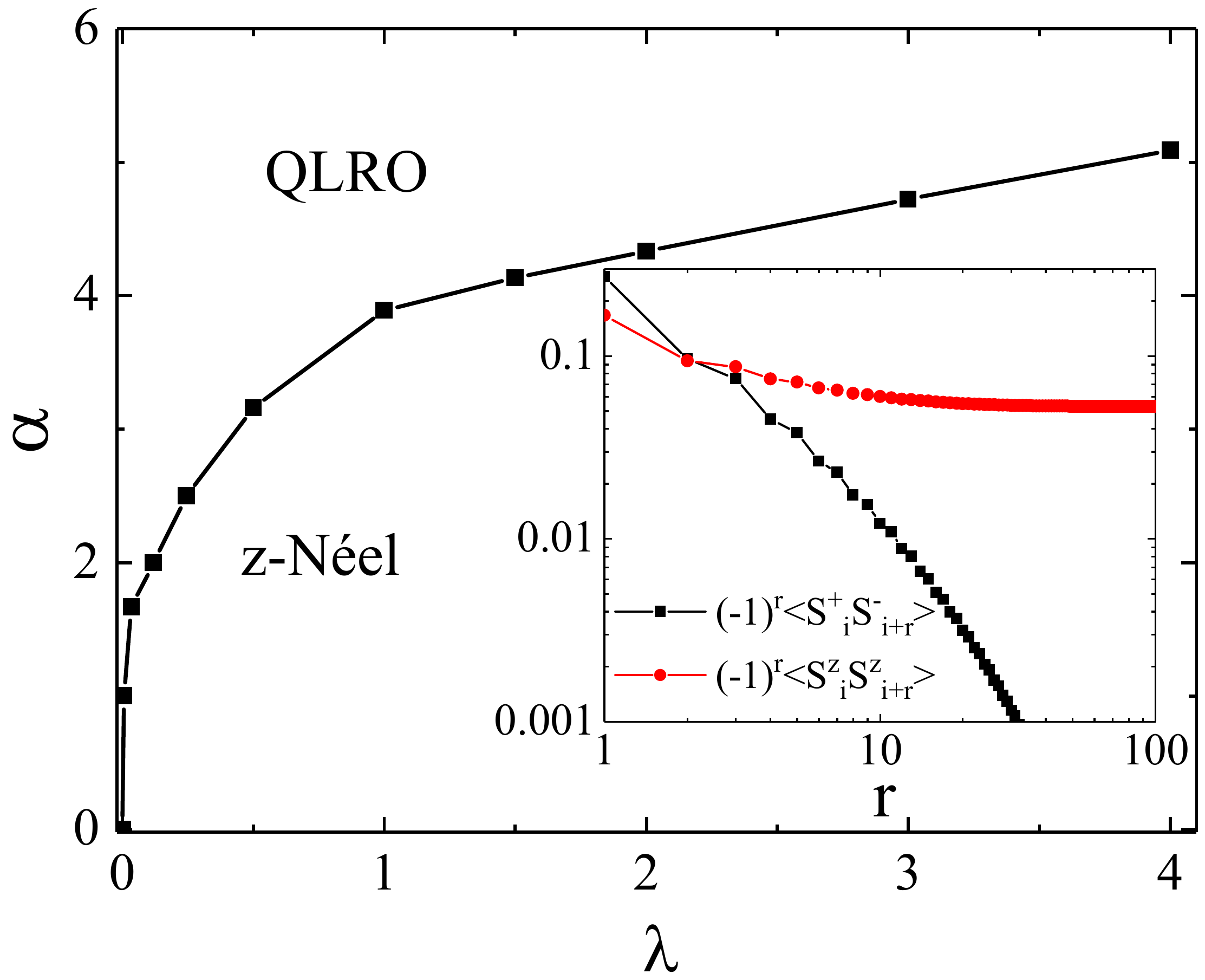}
\caption{\label{Phase_Ising} (a) Phase diagram of Hamiltonian Eq. (\ref{Hamiltonian}) as functions of $\alpha$ and $\lambda$ with $\Delta^{xy} =0$. Inset: Correlations $\langle S^{+}_iS^{-}_{i+r}\rangle$, $\langle S^{z}_iS^{z}_{i+r}\rangle$ versus the distance $r$ for $\lambda=1$, $\alpha=3$ with $\Delta^{xy} =0$.}
\end{figure}

\begin{figure}[t]
\includegraphics[width=0.50\textwidth]{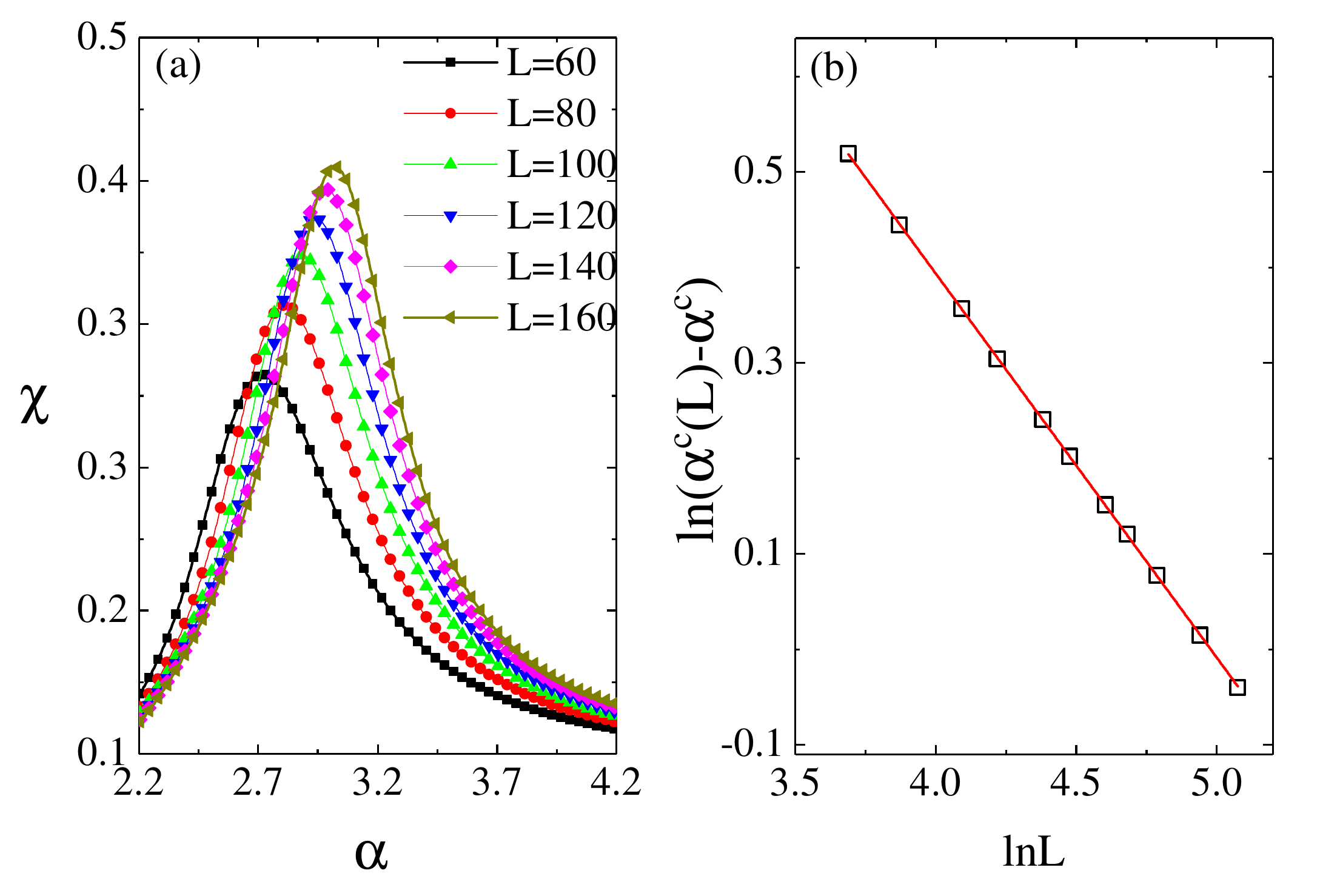}
\caption{(a) Fidelity susceptibility per site is plotted as a function of the parameter $\alpha$ on different system sizes $L$ with $\Delta^{xy} =0$, $\lambda=1$. (b) The corresponding scaling of the peak positions of $\chi$.}
\label{FS_Ising}
\end{figure}

\begin{figure}[t]
\includegraphics[width=0.5\textwidth]{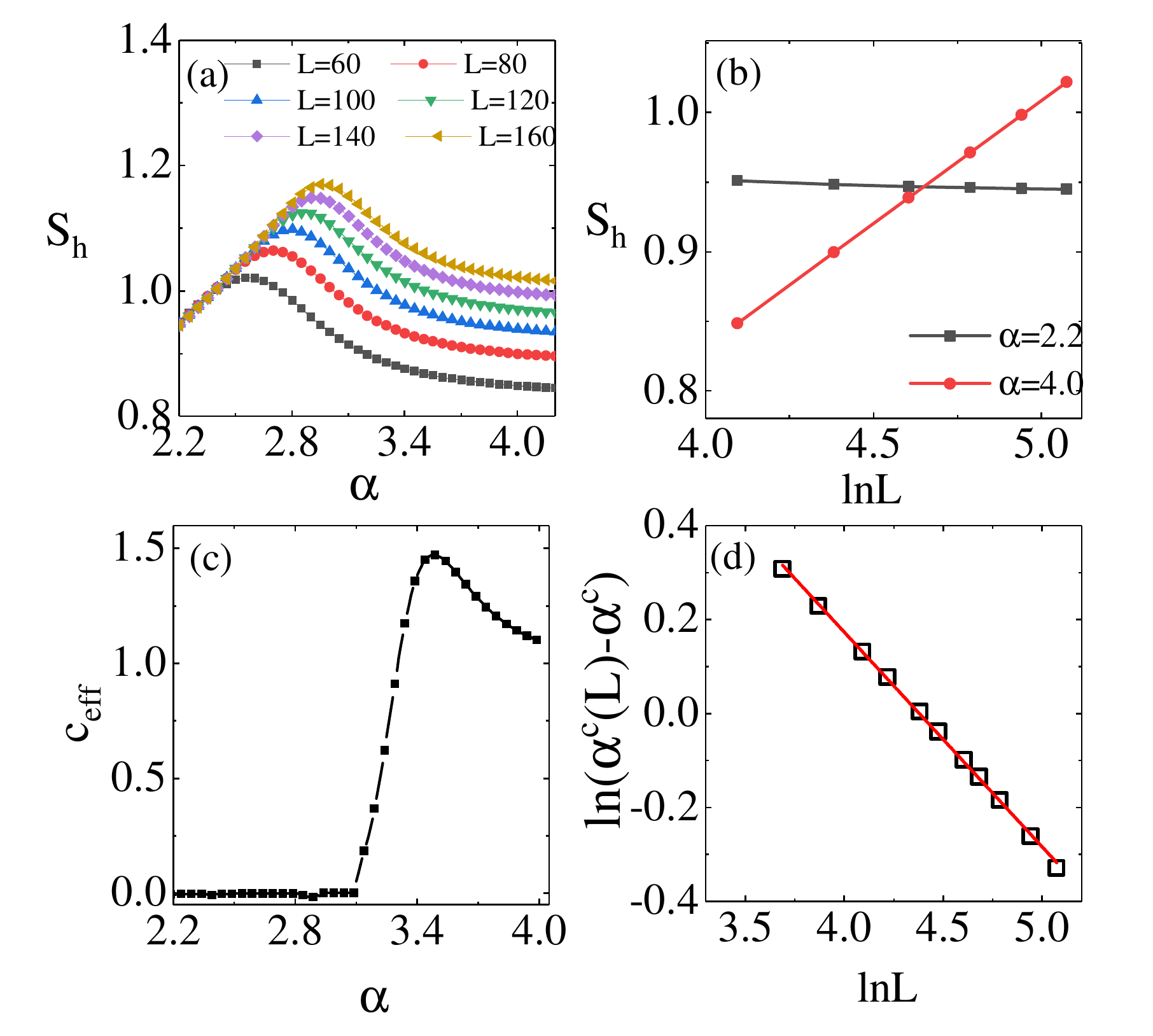}
\caption{(a)  Half-chain EE is plotted as a function of the decay exponent $\alpha$ on different system sizes $L$ with $\Delta^{xy} =0$, $\lambda=1$. (b) Half-chain EE versus $\ln L$ for different $\alpha$. (c) The effective central charge in Eq. (\ref{eq10}), versus $\alpha$ with $L_1=400$ and $L_2=500$. (d) The corresponding scaling of peak positions of $S_{\rm h}$.
\label{EE_Ising}}
\end{figure}

We can consider the Ising-type ($\Delta^{xy} <1$) long-range interactions. Here we exhibit a special case, i.e.,  $\Delta^{xy} =0$. The phase diagram of Hamiltonian Eq. (\ref{Hamiltonian}) as functions of $\alpha$ and $\lambda$ is shown in Fig.~\ref{Phase_Ising}. For sufficiently large $\alpha$, the system also would enter the QLRO phase. As the decay exponent $\alpha$ decreases, the long-range Ising interactions will become dominated. The correlation functions for $\{\alpha=3$, $\lambda=1\}$ are shown in the inset of Fig. \ref{Phase_Ising},
where $\langle S^+_{i}S^-_{i+r}\rangle$ exhibits an oscillating decay until vanishes as $r \to \infty$,
while $\langle S^z_{i}S^z_{i+r}\rangle$ 
alternates between $-0.053$ and $0.053$, implying
the characteristic of ${\mathbb Z}_2$ symmetry broken $z$-N\'eel phase.

Moreover, we find that the phase transitions between $z$-N\'{e}el and QLRO can be sensitively detected by both the FS and the EE.
In Fig. ~\ref{FS_Ising}(a),
the FS per site with respect to $\alpha$ for different system sizes $L$ is presented and
the peak of the ground-state FS becomes pronounced with increasing system sizes, which signals the occurrence of the QPT.
Regarding the finite-size scaling in Eq.(\ref{eq5}), $\alpha_c=3.88$ and $b=0.40$ can be extracted from Fig. \ref{FS_Ising}(b).
Further evidence for indicating the $z$-N\'{e}el-to-QLRO transition is provided by the EE, which is shown in Fig. ~\ref{EE_Ising}(a). Upon increasing the system size $L$, the EE shows a logarithmic growth for $\alpha>\alpha_c$ but saturates quickly otherwise [see  Fig.~\ref{EE_Ising}(b)], suggesting the $z$-N\'{e}el phase is gapped and the breaking of conformal symmetry. The effective central charge $c_{\rm eff}$ would  be zero [cf. \ref{EE_Ising}(c)].
Similar to that of the FS, the finite-size scaling of the EE yields $\alpha_c=3.88$,  $b=0.457$, as is exhibited in Fig. ~\ref{EE_Ising}(d).
It is worthy noting that the critical point $\alpha_c$ with $\Delta^{xy} =0$ becomes vanishing when the parameter $\lambda$ tends to zero, and diverges when $\lambda$ increases to the infinity.

\section{Discussion}

In this paper, we have studied the quantum phase transitions (QPTs) in the one-dimensional spin-$1/2$ chains with modulated long-range power-law-decaying interactions in terms of the density-matrix renormalization group technique. Together with the correlations and the entanglement entropy (EE), the ground-state fidelity susceptibility (FS) are employed to determine the phase boundary.
The XY-type long-range interactions lead to the emergence of U(1)-symmetric broken $xy$-N\'{e}el phase with long-range order (LRO) along easy axes~\cite{ZehanLi}, akin to the SU(2) symmetry broken N\'{e}el phase  induced by isotropic long-range interactions, while the Ising-type long-range interactions prompt the ${\mathbb Z}_2$ symmetry broken $z$-N\'{e}el phase. The FS can detect the QPT between the gapless quasi-long-range order (QLRO) phase and three different N\'{e}el phase, whether it is gapless or not. The FS proved to be a reliable tool to determine the ground-state phase diagram.  An area-law scaling is still valid in the gapped phase in the presence of the long-range interactions, although it was originally derived for the short-range interacting Hamiltonian. Figures \ref{EE_scaling}(a) and \ref{EE_Jxy}(a) demonstrate that the half-chain EE satisfies a logarithmic scaling with respect to the system size in gapless phases.  In this respect, the half-chain EE can faithfully seize the QPT between the gapless QLRO phase and the gapped $z$-N\'{e}el phase, while it is insensitive to QPTs between two gapless phases, such as QLRO to $xy$-N\'{e}el phase transition, QLRO to N\'{e}el phase transition. The insensitivity of the EE at quantum critical points between gapless phases
may be traced back to the gapless mode associated with the
spontaneous breaking of the continuous symmetry, sparking the challenge to demand much larger-scale computation for the effective central charge. In this context, using the maximum of bipartite EE as an indicator of a QPT from a gapless phase to another gapless phase is still elusive.  The models under consideration could be
envisioned in quantum simulation in
ultracold atoms~\cite{Periwal21,Takahashi}, opening the prospect for experimental investigation of the issues confronted here.
\\
\\
\begin{acknowledgments}
W.-L. You appreciates the valuable discussion with Gaoyong Sun and Ming Xue. This work is supported by the National Natural Science
Foundation of China (NSFC) under Grants No.~11104021, No.~12174194. J.R kindly acknowledge support from Open Project of  Key Laboratory of Artificial Structures and Quantum Control(Ministry of Education), Shanghai Jiao Tong University. W.-L. Y acknowledges the startup fund of Nanjing University of Aeronautics and Astronautics under Grant No. 1008-YAH20006, Top-notch Academic Programs Project of Jiangsu Higher Education Institutions and stable supports for basic institute research (Grant No. 190101).
\end{acknowledgments}

\end{document}